\documentclass[12pt]{article}

\usepackage{graphicx}
\usepackage{amsmath}

\begin{document}

\begin{center}
{\Large {\bf  Wigner representation for entanglement swapping using parametric down conversion:
the role of vacuum fluctuations in teleportation}}

\vspace{0.75cm}

{\bf A. Casado$^1$, S. Guerra$^2$, and J. Pl\'{a}cido$^3$}.

$^1$ Departamento de F\'{\i}sica Aplicada III, Escuela Superior de
Ingenieros,

Universidad de Sevilla, 41092 Sevilla, Spain.

Electronic address: acasado@us.es

$^2$ Centro Asociado de la Universidad Nacional de Educaci\'on a Distancia de Las Palmas de Gran Canaria,

 35004 Las Palmas de Gran Canaria, Spain.

$^3$ Grupo de Ingenier\'{\i}a T\'ermica e Instrumentaci\'on, Universidad de Las Palmas de Gran
Canaria,

 35017 Las Palmas de Gran Canaria, Spain.
\vspace{1cm}
\end{center}


\vspace{0.5cm}
\noindent {\bf Acknowledgements}
The authors would like to thank Prof. E. Santos for revising the
manuscript, and for helpful suggestions and comments on the work. A.
Casado acknowledges the support from the Spanish MCI Project no.
FIS2011-29400.

\vspace{0.5cm}
\noindent {\bf Abstract}

We apply the Wigner formalism of quantum optics in the Heisenberg
picture to study the role of the zeropoint field fluctuations in
entanglement swapping produced via parametric down conversion. It is
shown that the generation of mode entanglement between two initially
non interacting photons is related to the quadruple correlation
properties of the electromagnetic field, through the stochastic
properties of the vacuum. The relationship between the process of
transferring entanglement and the different zeropoint inputs at the
nonlinear crystal and the Bell-state analyser is emphasized.


\vspace{1cm}

\newpage
\section{Introduction}
The theory of quantum information in quantum optics is supported by the
phenomena of entanglement \cite{epr, ent1} and hyperentanglement
\cite{hip, hip1, hip2} produced via parametric down conversion (PDC)
\cite{pdc, swp6, pdc1, pdc2}. These phenomena constitute a very
important experimental arena for quantum cryptography \cite{ekert,
gisin}, dense coding \cite{dc}, superdense coding \cite{dc1} and
teleportation \cite{nswp, swp9, swp10, tele}. The ultimate goal would
be to build a quantum computer network for the transmission and
reconstruction over an arbitrary distance of a quantum state, but for
the latter, it would be necessary to build a network of repeaters
\cite{swpG, swpH, swpF, swpD, rep, rep1} whose physical base is the
process known as entanglement swapping \cite{swp}. This implies that
two particles which have never interacted are entangled as a
consequence of a Bell state measurement (BSM) \cite{swp14,swpC},
involving two Einstein-Podolsky-Rosen (EPR) pairs \cite{epr}. In 2000,
Asher Peres \cite{swpZ} put forward the paradoxical idea that
entanglement could be produced after the entangled particles have been
measured, even if they no longer exist. This can also be viewed as
quantum steering into the past. Recent studies appear to confirm this
paradox \cite{swpB}. More recently, it has been demonstrated that
entanglement can be transferred to two photons that exist at separate
times \cite{nueva2013}.

The Wigner formalism constitutes a complementary approach to the
standard Hilbert-space formulation for the study of optical quantum
information processing and for its practical implementation using PDC.
In the Wigner representation within the Heisenberg picture (WRHP), the
Wigner function is time independent, it corresponds to the Wigner
distribution of the initial state of the electromagnetic field, and the
dynamics is contained in the electric field amplitudes. The analysis of
the generation and propagation of PDC light with this formalism was
treated in a series of papers using a Hamiltonian approach \cite{swpL,
swpM} and also by starting from the Maxwell equations inside the
crystal \cite{swp19}. The WRHP approach resembles classical optics, in
the sense that the light emitted by the crystal is generated via the
coupling between the zeropoint field (ZPF) and the laser beam entering
the nonlinear medium, which gives rise to an amplification of vacuum
fluctuations. The Wigner function is positive in this case, it
corresponds to the Gaussian Wigner distribution of the vacuum
amplitudes. Finally, the zeropoint fluctuations are subtracted at the
detectors, and the theory of detection in the Wigner approach shows how
the signal is separated from the ZPF background. These two features,
zeropoint field and its subtraction in the detection process,
constitute the main differences with respect to classical optics, and
give rise to the typical results within the quantum domain.

The WRHP formalism has been applied recently to the study of
experiments on quantum communication using PDC. For instance, quantum
cryptography with entangled photons \cite{swpI}, partial Bell-state
measurement \cite{swpJ}, and polarization-momentum hyperentanglement
and its application to complete Bell-state measurement \cite{swpK}. The
essential point of the WRHP formalism is that it focuses on the
relationship between the correlation properties of the light field in a
concrete experiment, through the propagation of the zeropoint field
amplitudes, and the corresponding optical quantum communication
protocol. As a matter of fact, a key point of this approach is that
entanglement can be seen just as an interplay of correlated waves,
which sharply contrasts to the usual Hilbert-space or more
particle-based formalism \cite{swpL}. Also there is a double role of
the zeropoint field in this kind of experiments: it carries the quantum
information that is extracted at the source and introduces a
fundamental noise at the idle channels of the analysers, which limits
the information that can be efectively measured \cite{swpK}. Hence, the
WRHP formalism offers a complementary wave-like reinterpretation of
experiments on quantum communication involving PDC light, to the one
provided by the standard Hilbert-space description based on the concept
of qubit, in which the corpuscular aspect of light is emphasized.

In this paper we shall analyze the relationship between entanglement
swapping using PDC and the zeropoint field fluctuations by using the
WRHP formalism. As we shall show throughout this document, the
fundamental concept for understanding the phenomenon of teleportation
of entanglement, is the quadruple correlation of the electromagnetic
field. Our analysis using the WRHP approach will contrast, apparently,
to the usual explanation in terms of the collapse of the state vector
at the Bell-state analyser. The link between both formalisms is found
to be in the zeropoint field amplitudes, but the importance of this
work goes beyond the mathematical aspects, giving rise to new results.

This paper is organised as follows: In Section \ref{sec2} we shall give
the WRHP description of the basic quantum state for entanglement
swapping \cite{swp2}, and we shall calculate the field amplitudes at
the detectors. We shall show that, in order to generate the two pair of
entangled beams, eight sets of independent zeropoint modes (four sets
for each emission) are necessary. In Subsection \ref{crosco} we shall
calculate the cross-correlation properties of the light field. In
Section \ref{sec3} the quadruple correlations leading to four
fold-coincidence will be analysed, and the intrinsic nature of
teleportation based in the zeropoint field amplitudes will be revealed.
Finally, in Section \ref{sec5} we shall discuss the main results of
this work, and we shall present the conclussions and further steps of
this research line. In Appendix \ref{A} we present a brief summary of
the most important concepts and results of the WRHP formalism of PDC,
which are used in this paper. In addition, we have included in Appendix
\ref{B} the calculation of the quadruple detection probability in the
WRHP approach.

\section{Entanglement Swapping in the WRHP formalism}
\label{sec2}
\begin{figure}[h]
      \centering
      \includegraphics[width=9.0cm,clip]{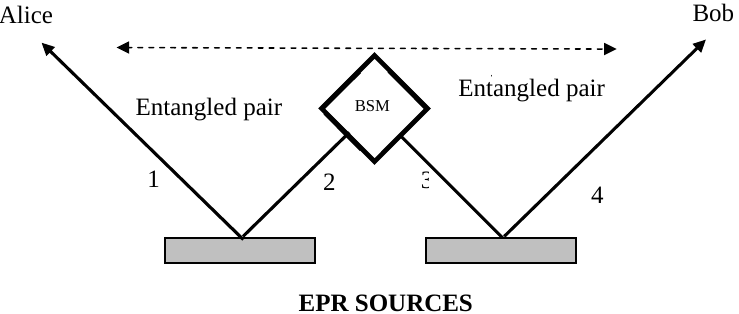}
\caption{\small{Principle of entanglement swapping.}}
\label{Figure1}
\end{figure}

Entanglement swapping \cite{swp, swp4, swp5, swp3, swp1} provides a method of entanglement of two particles that do not interact. Let us review the basic aspects of this process in the Hilbert space \cite{nueva1}. 
Two EPR sources generate, independently, two pairs of entangled
particles, pair $1$-$2$ and pair $3$-$4$, being each pair described by a singlet
state. Particles $2$ and $3$ are subjected to a Bell state analysis as shown
in Figure \ref{Figure1}. The collapse of the state vector to a given eigenstate of particles $2$ and $3$ gives rise to an entanglement between particles $1$ and $4$, which
is called entanglement teleportation or entanglement swapping.

The total state describes the fact that particles $1$ and $2$ ($3$ and $4$) are entangled
in a singlet state. For instante, if we are dealing with polarization entanglement, we have:

\[
|\Pi \rangle_{1234} =| \Psi ^{-}  \rangle_{12} | \Psi ^{-}
\rangle_{34}
\]
\begin{equation}
=\frac{1}{2} (|\Psi^{+}\rangle_{14}|\Psi^{+}\rangle_{23}-|\Psi ^{-}
\rangle_{14}| \Psi ^{-}\rangle_{23}-| \Phi ^{+}\rangle_{14}| \Phi ^{+}
\rangle_{23} +| \Phi ^{-}\rangle_{14} |\Phi ^{-} \rangle_{23}),
\label{40}\end{equation}
where

\begin{equation}| \Psi ^{\pm}\rangle_{ij}  =\frac{1}{\sqrt{2}}
(| H\rangle_{i}| V\rangle_{j}
\pm | V \rangle_{i}| H \rangle_{j})
\,\,\,;\,\,\,| \Phi ^{\pm}\rangle_{ij} =\frac{1}{\sqrt{2}}
(| H\rangle_{i}| H\rangle_{j}
\pm | V \rangle_{i}| V \rangle_{j})
,\label{s1}
\end{equation}
are the polarization Bell base states. Factoring is a consequence of
the pairs being independent. Nevertheless, a BSM on particles $2$ and
$3$ will leave particles $1$ and $4$ entangled in the same state as the
corresponding to the projective measurement on the pair $2-3$. In this
way, particles $1$ and $4$ will end up in one of the four Bell states:
$|\Psi_{14}^{+}\rangle$, $|\Psi_{14}^{-}\rangle$,
$|\Phi_{14}^{+}\rangle$ and $|\Phi_{14}^{-}\rangle$, with the same $1/4$ probability.
In the case of light, it is well known that the four Bell states are not distinguishable
when entanglement in only one degree of freedom is considered \cite{hp}. In the last decade,
the problem of performing complete Bell-state measurement with photons has been solved using
hyperentanglement \cite{hip2}. 

\begin{figure}[h]
      \centering
      \includegraphics[width=10.0cm,clip]{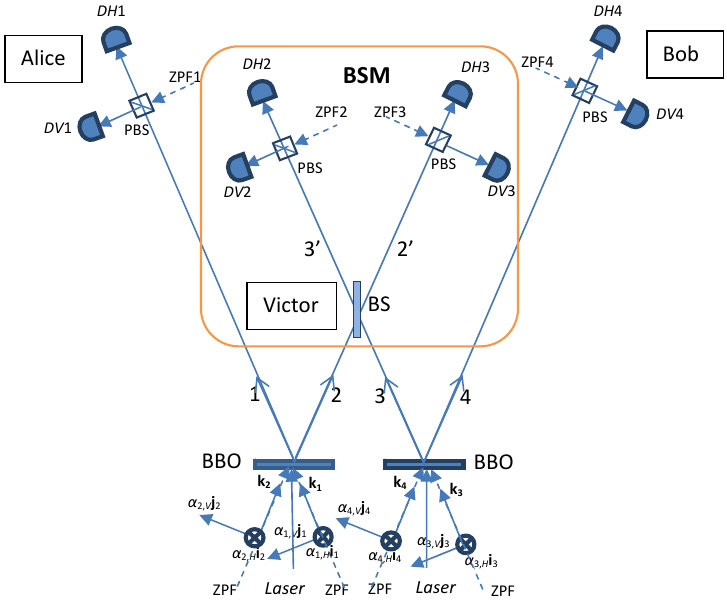}
\caption{\small{Entanglement swapping via two nonlinear crystals. Each couple of down converted photons is generated through the coupling inside the nonlinear source between the laser and four sets of independent zeropoint modes. This is an essential aspect in the WRHP description for PDC experiments.}}
\label{Figure2}
\end{figure}

The quantum predictions corresponding to the state $| \Pi
\rangle_{1234}=|\Psi^{-}\rangle_{12} |\Psi ^{-}  \rangle_{34}$ are
reproduced in the WRHP approach (see Appendix \ref{A}) via the
consideration of the following four beams (see Fig. \ref{Figure2}),
which are generated from the coupling inside the nonlinear medium
between the laser field and the ZPF inputs (see Eqs. (\ref{hyper0}) and
(\ref{hyper3})):

\begin{equation}\begin{array}{l}{\bf{F}}_{1}^{(+)}({\bf r}_{S_1}, t)=
{F'}_{s}^{(+)}({\bf r}_{S_1}, t ;\{\alpha _{{\bf{k}}_{1},H}; \alpha^{*} _{{\bf{k}}_{2}, V}\}){\bf{i}}_{1}+
{F'}_{p}^{(+)}({\bf r}_{S_1}, t; \{\alpha _{{\bf{k}}_{1},V}; \alpha^{*}_{{\bf{k}}_{2}, H}\}){\bf{j}}_{1},
\end{array}\label{GrindEQ__6_3_}\end{equation}
\begin{equation}\begin{array}{l}{\bf{F}}_{2}^{(+)}({\bf r}_{S_1}, t)=
{F'}_{q}^{(+)}({\bf r}_{S_1}, t ;\{\alpha _{{\bf{k}}_{2},H}; \alpha^{*} _{{\bf{k}}_{1}, V}\}){\bf{i}}_{2}-
{F'}_{r}^{(+)}({\bf r}_{S_1}, t; \{\alpha _{{\bf{k}}_{2},V}; \alpha^{*}_{{\bf{k}}_{1}, H}\}){\bf{j}}_{2},
\end{array}\label{GrindEQ__6_3_bis}\end{equation}
\begin{equation}\begin{array}{l}{\bf{F}}_{3}^{(+)}({\bf r}_{S_2}, t)=
F_{s}^{(+)}({\bf r}_{S_2}, t; \{\alpha _{{\bf{k}}_{3},H};
\alpha^{*}_{{\bf{k}}_{4},V}\}){\bf{i}}_{3}+ F_{p}^{(+)}({\bf r}_{S_2},
t; \{\alpha _{{\bf{k}}_{3},V};\alpha^{*}
_{{\bf{k}}_{4},H}\}){\bf{j}}_{3},
\end{array}\label{GrindEQ__6_6_bis}\end{equation}
\begin{equation}\begin{array}{l}{\bf{F}}_{4}^{(+)}({\bf r}_{S_2} ,t)=
F_{q}^{(+)}({\bf r}_{S_2}, t; \{\alpha _{{\bf{k}}_{4},H};\alpha^{*}
_{{\bf{k}}_{3},V}\}){\bf{i}}_{4}- F_{r}^{(+)}({\bf r}_{S_2} , t;
\{\alpha _{{\bf{k}}_{4},V};\alpha^{*} _{{\bf{k}}_{3},
V}\}){\bf{j}}_{4},
\end{array}\label{GrindEQ__6_6_}\end{equation}
where we have considered that the center of the first (second)
nonlinear source is located at position ${\bf r}_{S_1}$ (${\bf
r}_{S_2}$). We have included the sets of zeropoint modes that appear in
each electric field component \cite{swpJ}. The only non-null
cross-correlations are those concerning the labels $(p, q)$ and $(r,
s)$ of beams $1-2$, and the same for the beams $3-4$, i.e. the non-null
cross-correlations correspond to different polarization components, and
there is a sign difference between the two correlations involving the
couple of beams $1-2$ ($3-4$), as it can be seen from Eqs.
\eqref{GrindEQ__6_3_} and \eqref{GrindEQ__6_3_bis}
[\eqref{GrindEQ__6_6_bis} and \eqref{GrindEQ__6_6_}]. On the other
hand, there is no cross-correlation involving any of the primed
amplitudes (beams $1-2$) with the unprimed ones (beams $3-4$).

Taking into account the propagation of the field amplitudes [see Eq.
(\ref{propagation})] we can express all the cross-correlations, at any
position and time, in terms of the corresponding ones at the center of
the nonlinear sources \cite{swpM}. For instance (see Eq. (\ref{nu})):

\begin{equation}
\langle F^{(+)}_{p}({\bf r}_{S_2}, t)F^{(+)}_{q}({\bf r}_{S_2}, t')\rangle
=gV\nu(t'-t).
\label{nu}
\end{equation}
A similar expression holds for $\langle F^{(+)}_{r}({\bf r}_{S_2},
t)F^{(+)}_{s}({\bf r}_{S_2}, t')\rangle$, and also for the
corresponding primed amplitudes, which are emitted at the first
crystal.

Let us emphasize that beams $1$ and $2$ are completely uncorrelated
with  beams  $3$ and $4$, given that the vacuum modes that are involved
in the couple $1-2$ share no correlation with the corresponding vacuum
modes in the couple $3-4$, as it can be easily seen from Eq.
(\ref{correlations}). This is closely related to the factorization in
Eq. (\ref{40}).

Where is the origin of entanglement swapping in the WRHP formalism? The answer to this question is centralised in how the correlations change, after beams
$2$ and $3$ cross the balanced beam-splitter (BS). The action of the BS on beams $2$ and $3$ produces beams ${\bf{F}}_{3'}^{(+)}$ and ${\bf{F}}_{2'}^{(+)}$:
\[
{\bf{F}}_{3'}^{(+)}({\bf r}_{BS}, t)=\frac{1}{\sqrt{2}}[i{F'}_{q}^{(+)}({\bf r}_{BS}, t)
+F_{p}^{(+)}({\bf r}_{BS}, t)]{\bf{i}}_{3'}
\]
\begin{equation}
+\frac{1}{\sqrt{2}}[i{F'}_{r}^{(+)}({\bf r}_{BS}, t)+F_{s}^{(+)}({\bf
r}_{BS}, t)]{\bf{j}}_{3'},
\label{bs1}
\end{equation}

\[
{\bf{F}}_{2'}^{(+)}({\bf r}_{BS}, t)=\frac{1}{\sqrt{2}}[{F'}_{q}^{(+)}({\bf r}_{BS}, t)
+iF_{p}^{(+)}({\bf r}_{BS}, t)]{\bf{i}}_{2'}
\]
\begin{equation}
+\frac{1}{\sqrt{2}}[{F'}_{r}^{(+)}({\bf r}_{BS}, t)+iF_{s}^{(+)}({\bf
r}_{BS}, t)]{\bf{j'}}_{2'},
\label{bs2}
\end{equation}
with ${\bf r}_{BS}$ being the position of the beam-splitter where the
two beams are recombined. Now, we shall consider that the polarization
beam splitters (PBS) transmit (reflect) horizontal (vertical)
polarization. The electric field amplitude at the detectors will
include the superposition of a zeropoint field component coming from
the idle channels of the PBSs. We shall use unprimed (primed)
space-time variables for characterizing the field amplitude at the
detector corresponding to horizontal (vertical) polarization:

\begin{equation}
({\bf r}_{DHi}, t_{DHi}) \equiv ({\bf r}_{i}, t_i)
\,\,\,\,;\,\,\,\,
({\bf r}_{DVi}, t_{DVi}) \equiv ({\bf r'}_{i},
t'_i)\,\,\,\,;\,\,\,\,i=1, 2, 3, 4.
\end{equation}
In this way, the field amplitudes at the Bell-state analyser are:

\begin{equation}\begin{array}{l}{\bf{F}}_{DH2}^{(+)}({\bf{r}}_{2} , t_{2})=
\frac{1}{\sqrt{2}}[i{F'}_{q}^{(+)}({\bf{r}}_{2} , t_{2})+F_{s}^{(+)}({\bf{r}}_{2} ,t_{2} )]{\bf{i}}+i[{\bf{F}}_{ZPF2}^{(+)} ({\bf{r}}_{2} , t_{2})\cdot {\bf{j}}]{\bf{j}},\end{array}\label{1000}\end{equation}

\begin{equation}\begin{array}{l}{\bf{F}}_{DV2}^{(+)} ({\bf{r'}}_{2} ,t'_{2} )=\frac{i}{\sqrt{2} } [i{F'}_{r}^{(+)}({\bf{r'}}_{2} , t'_{2})-F_{p}^{(+)} ({\bf{r'}}_{2} ,t'_{2})]{\bf{j}}+[{\bf{F}}_{ZPF2}^{(+)}({\bf{r'}}_{2} ,t'_{2})\cdot {\bf{i}}]{\bf{i}},\end{array}\label{60}\end{equation}

\begin{equation}\begin{array}{l}{\bf{F}}_{DH3}^{(+)}({\bf{r}}_{3} , t_{3})=
\frac{1}{\sqrt{2}}[{F'}_{q}^{(+)}({\bf{r}}_{3} , t_{3})+iF_{s}^{(+)}({\bf{r}}_{3} , t_{3})]{\bf{i'}}+i[{\bf{F}}_{ZPF3}^{(+)}({\bf{r}}_{3} , t_{3})\cdot {\bf{j'}}]{\bf{j'}},\end{array}
\label{60bis}
\end{equation}

\begin{equation}\begin{array}{l}{\bf{F}}_{DV3}^{(+)}({\bf{r'}}_{3}, t'_{3})=\frac{i}{\sqrt{2}}[{F'}_{r}^{(+)}({\bf{r'}}_{3}, t'_{3})-iF_{p}^{(+)}({\bf{r'}}_{3}, t'_{3} )]{\bf{j'}}+[{\bf{F}}_{ZPF3}^{(+)}({\bf{r'}}_{3}, t'_{3})\cdot {\bf{i}}']{\bf{i'}},\end{array}\label{61}\end{equation}
On the other hand, the detector outputs which are located in beams $1$
and $4$, are:

\begin{equation}\begin{array}{l}{\bf{F}}_{DH1}^{(+)}({\bf{r}}_{1} , t_{1})={F'}_{s}^{(+)} ({\bf{r}}_{1} ,t_{1} ){\bf{i}}+i[{\bf{F}}_{ZPF1}^{(+)}({\bf{r}}_{1} , t_{1})\cdot {\bf{j}}]{\bf{j}}, \end{array}\label{62}\end{equation}
\begin{equation}\begin{array}{l}{\bf{F}}_{DV1}^{(+)}({\bf{r'}}_{1} , t'_{1})=-i{F'}_{p}^{(+)} ({\bf{r'}}_{1} ,t'_{1} ){\bf{j}}+[{\bf{F}}_{ZPF1}^{(+)} ({\bf{r'}}_{1} , t'_{1})\cdot {\bf{i}}]{\bf{i}},\end{array}\label{63}\end{equation}

\begin{equation}\begin{array}{l}{\bf{F}}_{DH4}^{(+)}({\bf{r}}_{4} , t_{4})=F_{q}^{(+)}({\bf{r}}_{4} ,t_{4} ){\bf{i'}}+i[{\bf{F}}_{ZPF4}^{(+)}({\bf{r}}_{4} , t_{4})\cdot {\bf{j'}}]{\bf{j'}},\end{array}\label{56}\end{equation}
\begin{equation}\begin{array}{l}{\bf{F}}_{DV4}^{(+)}({\bf{r'}}_{4} , t'_{4})=iF_{r}^{(+)}({\bf{r'}}_{4} , t'_{4}){\bf{j'}}+[{\bf{F}}_{ZPF4}^{(+)}({\bf{r'}}_{4} , t'_{4})\cdot {\bf{i'}}]{\bf{i'}}. \end{array}\label{57}\end{equation}


The beams given in Eqs.
\eqref{GrindEQ__6_3_}, \eqref{GrindEQ__6_3_bis},
\eqref{GrindEQ__6_6_bis} and \eqref{GrindEQ__6_6_} allow for a
wave-like description which reproduces the same results that the
corresponding to the use of (\ref{40}). Nevertheless, one of the critical matters related to
this description is that PDC light generally have higher-order
correlated photons and the assumption of having the singlet state is
very limited. For low values of the intensity
of the laser only one pair of photons is generated with a probability
much lower than unity, so that a higher photon number of contributions
are insignificant.  When increasing the intensity of the laser, higher pair
generation rates are possible, but also higher order components are
increased \cite{higher}. Nevertheless, the WRHP approach can be adequate to
take into consideration higher-order processes into the crystal: all
the information is included in the polarization components of the
field, which could be calculated to higher orders in the coupling
constant, allowing for a description of higher-order correlated
photons.

\subsection{Cross-correlations}
\label{crosco}
Now, let us study the cross-correlation properties of the light field.
Each detector is reached by the two polarization components, one of
them being zeropoint radiation which is transmitted (reflected) at the
vertical (horizontal) outgoing channel of the corresponding PBS, which
does not correlate with any other amplitude. In this way, the field
amplitude at each detector has two components, $x\equiv horizontal$,
$y\equiv vertical$, so that:

\begin{equation}
{\bf F}_{DHi}^{(+)}=F_{DHi, x}^{(+)}{\bf i}_{i}+F_{DHi, y}^{(+)}{\bf
j}_{i}\,\,\,\,;\,\,\,\,{\bf F}_{DVi}^{(+)}=F_{DVi, x}^{(+)}{\bf
i}_{i}+F_{DVi, y}^{(+)}{\bf j}_{i}\,\,\,\,;\,\,\,\,i=\{1, 2, 3, 4\}.
\end{equation}
For notation simplicity, we shall make the change:

\begin{equation}
F_{z}^{(+)}({\bf r}_{i}, t_{i})\equiv F_{z,
i}^{(+)}\,\,\,\,;\,\,\,\,F_{z}^{(+)}({\bf r'}_{i}, t'_{\i})\equiv F_{z,
i'}^{(+)}\,\,\,\,;\,\,\,\,z=\{p, q, r, s\}\,\,;\,\,i=\{1, 2, 3, 4\},
\end{equation}
and the same holds for the primed amplitudes.

Taking into consideration that there is no cross-correlation linked to
the couple of beams $1-4$, nor to the couple $2-3$, and that there is
no cross-correlation concerning two amplitudes with the same
polarization, there are only eight pairs of correlated amplitudes. The
effect of the BS is to duplicate the number of cross-correlations
corresponding to the outgoing beams given in Eqs. \eqref{GrindEQ__6_3_}
to \eqref{GrindEQ__6_6_}. We have:

\begin{itemize}

\item ${\bf{F}}_{DH2}^{(+)}$ is correlated to ${\bf{F}}_{DV1}^{(+)}$ and ${\bf{F}}_{DV4}^{(+)}$:

\begin{equation}
\langle F_{DH2,x}^{(+)}F_{DV1,y}^{(+)} \rangle =\frac{1}{\sqrt{2}}\langle {F'}_{q, 2}^{(+)}{F'}_{p, 1'}^{(+)}\rangle
\,\,\,\,;\,\,\,\,
\langle F_{DH2,x}^{(+)}F_{DV4,y}^{(+)} \rangle =\frac{i}{\sqrt{2} }\langle F_{s, 2}^{(+)} F_{r, 4'}^{(+)}\rangle.
\label{22primera}
\end{equation}

\item ${\bf{F}}_{DV2}^{(+)}$ is corretaled to ${\bf{F}}_{DH1}^{(+)}$ and ${\bf{F}}_{DH4}^{(+)}$:

\begin{equation}
\langle F_{DV2,y}^{(+)} F_{DH1,x}^{(+)}\rangle =\frac{-1}{\sqrt{2}}\langle {F'}_{r, 2'}^{(+)}{F'}_{s, 1}^{(+)}\rangle
\,\,\,\,;\,\,\,\,
\langle F_{DV2,y}^{(+)} F_{DH4,x}^{(+)}\rangle =\frac{-i}{\sqrt{2} }\langle F_{p, 2'}^{(+)} F_{q, 4}^{(+)}\rangle.
\label{22segunda}
\end{equation}

\item ${\bf{F}}_{DH3}^{(+)}$ is correlated to ${\bf{F}}_{DV1}^{(+)}$ and ${\bf{F}}_{DV4}^{(+)}$:

\begin{equation}
\langle F_{DH3,x}^{(+)}F_{DV1,y}^{(+)}\rangle =\frac{-i}{\sqrt{2}}\langle {F'}_{q, 3}^{(+)} {F'}_{p, 1'}^{(+)}\rangle
\,\,\,\,;\,\,\,\,
\langle F_{DH3,x}^{(+)}F_{DV4, y}^{(+)} \rangle =\frac{-1}{\sqrt{2}}\langle F_{s, 3}^{(+)} F_{r, 4'}^{(+)}\rangle.
\label{22tercera}
\end{equation}

\item ${\bf{F}}_{DV3}^{(+)}$ is correlated to ${\bf{F}}_{DH1}^{(+)}$ and ${\bf{F}}_{DH4}^{(+)}$:

\begin{equation}
\langle F_{DV3,y}^{(+)}F_{DH1,x}^{(+)}\rangle =\frac{i}{\sqrt{2}}\langle {F'}_{r, 3'}^{(+)}{F'}_{s, 1}^{(+)}\rangle
\,\,\,\,;\,\,\,\,
\langle F_{DV3,y}^{(+)} F_{DH4,x}^{(+)}\rangle =\frac{1}{\sqrt{2}}\langle F_{p, 3'}^{(+)} F_{q, 4}^{(+)} \rangle. \label{22}\end{equation}

\end{itemize}

The values of the above cross-correlations depend on the values of the space-time variables related to each detector amplitude, as it can be seen from Eqs. (\ref{propagation}) and (\ref{nu}).
For instance, by considering $t_i=t'_j$ and that there is an identical distance from each source to the detectors, the following result is obtained (see Eq. \eqref{p12}):
\begin{equation}
\frac{P_{DHi,DVj}}{K_{DHi}K_{DVj}}=\frac{g^2|V|^2}{2}\left|\nu(0)\right|^{2}
\,\,\,;\,\,\,i \neq j\,\,\,;\,\,\,(i,j)\neq (1,4), (2,3),
\label{probid}
\end{equation}
where $K_{DHi}$ and $K_{DVj}$ are constants related to the detection efficiency.


\section{Quadruple correlations}
\label{sec3}
The aim of this section is the understanding of the physics of
entanglement swapping, through the calculation of the quadruple
correlation properties of the electric field. The quadruple detection
probability [see Eq. (\ref{e10})] is expressed in terms of quadruple
correlations of the type $\langle {F}_{a, \lambda }^{(+)} {F}_{b,
\lambda' }^{(+)} {F}_{c, \lambda'' }^{(+)} {F}_{d, \lambda'''
}^{(+)}\rangle$. Taking into account that we are dealing with a
Gaussian process, and using Eq. (\ref{12}), we have

\begin{equation}
\begin{array}{l}
{\langle {F}_{a,\lambda }^{(+)} {F}_{b,\lambda' }^{(+)} {F}_{c,\lambda'' }^{(+)}{F}_{d,\lambda''' }^{(+)}\rangle
=\langle {F}_{a,\lambda }^{(+)} {F}_{b,\lambda' }^{(+)} \rangle \langle
{F}_{c,\lambda'' }^{(+)}
{F}_{d,\lambda''' }^{(+)} \rangle} \\
{\, \, \, \, \, \, \, \, \, \, \, \, \, \, \, \, \, \, \, \, \, \, \, \, \, \, \, \, \, \, \, \, \, \, \, \,
\, \, \, \, \, \, \, \, \, \, \, \, \, \, \, \, \, \, \, \, \, \, \, \, \, \, \,
+\langle {F}_{a,\lambda }^{(+)}{F}_{c,\lambda'' }^{(+)}\rangle  \langle
{F}_{b,\lambda' }^{(+)}
{F}_{d,\lambda'''}^{(+)} \rangle  +\langle {F}_{a,\lambda }^{(+)} {F}_{d,\lambda''' }^{(+)} \rangle
\langle {F}_{b,\lambda'}^{(+)} {F}_{c,\lambda''}^{(+)} \rangle}.\end{array}\label{12bis}\end{equation}

Let us study the situation described in figure \ref{Figure2}, and we
shall consider that \textit{a} (\textit{d}) is a label for a given
detector in beam area $1$ ($4$), and that $b$ and $c$ are referred to
the detectors at the Bell-state analyser. On the other hand, $\lambda$,
$\lambda'$, $\lambda''$ and $\lambda'''$ will label the corresponding
polarization of the field amplitude at the detector, i.e. $\{\lambda,
\lambda', \lambda'', \lambda'''\}=\{x ,y\}$. Because of beams $1$ and $4$
are uncorrelated, as well as $2$ and $3$, we can observe that the last
addend of
\eqref{12bis} is zero, so that
\begin{equation}
\langle {F}_{a,\lambda }^{(+)} {F}_{b,\lambda {'} }^{(+)} {F}_{c,\lambda {'} {'} }^{(+)}
{F}_{d,\lambda {'} {'} {'} }^{(+)} \rangle =\langle {F}_{a,\lambda }^{(+)} {F}_{b,\lambda {'} }^{(+)}
\rangle \langle {F}_{c,\lambda {'} {'} }^{(+)} {F}_{d,\lambda {'} {'} {'} }^{(+)} \rangle
+\langle {F}_{a,\lambda }^{(+)} {F}_{c,\lambda {'} {'} }^{(+)} \rangle
\langle {F}_{b,\lambda {'} }^{(+)} {F}_{d,\lambda {'} {'} {'} }^{(+)}
\rangle ,\label{20}\end{equation} which shows that the quadruple
correlation is generally different from zero, even if there are two
pairs of detector amplitudes which are uncorrelated.

Now we shall analyse, for each possible joint detection concerning the BSM of photons $2$ and $3$, the associated quadruple correlations:

\begin{enumerate}
\item Let us first analyse
the eight quadruple correlations in which both detectors of the same area, $DH2$ and $DV2$, or $DH3$ and $DV3$, are involved \cite{swp11}.
By using Eq. \eqref{20}, and taking into consideration that the cross-correlations corresponding to the same polarization are zero, only four correlations are different from zero, those concerning different polarization at detectors in beam areas $1$ and $4$. We have, for $i=2, 3$:


\begin{equation}
\langle {F}_{DH1,x}^{(+)} {F}_{DHi,x}^{(+)} {F}_{DVi,y}^{(+)} {F}_{DV4,y}^{(+)}\rangle
=-\frac{i}{2} \langle {F'}_{s, 1}^{(+)}{F'}_{r, i'}^{(+)}\rangle
\langle {F}_{s, i}^{(+)}{F}_{r, 4'}^{(+)}  \rangle,
\label{13}\end{equation}

\begin{equation}
\langle {F}_{DV1,x}^{(+)} {F}_{DHi,x}^{(+)} {F}_{DVi,y}^{(+)} {F}_{DH4,y}^{(+)}\rangle
=-\frac{i}{2} \langle {F'}_{p, 1'}^{(+)}{F'}_{q, i}^{(+)}\rangle
\langle {F}_{p, i'}^{(+)}{F}_{q, 4}^{(+)}  \rangle.
\label{14}\end{equation}


\item Now we shall
study the eight quadruple correlations in which the detectors $DH2$ and $DV3$, or $DV2$ and $DH3$, are involved.
By using Eq. \eqref{20}, only four correlations are different from zero:



\begin{equation}
\langle {F}_{DH1,x}^{(+)} {F}_{DH2,x}^{(+)} {F}_{DV3,y}^{(+)} {F}_{DV4,y}^{(+)}\rangle
=-\frac{1}{2} \langle {F'}_{s, 1}^{(+)}{F'}_{r, 3'}^{(+)}\rangle
\langle {F}_{s, 2}^{(+)}{F}_{r, 4'}^{(+)}  \rangle,
\label{15bb}
\end{equation}
\begin{equation}
\langle {F}_{DH1,x}^{(+)} {F}_{DV2,y}^{(+)} {F}_{DH3,x}^{(+)} {F}_{DV4,y}^{(+)}\rangle
=+\frac{1}{2} \langle {F'}_{s, 1}^{(+)}{F'}_{r, 2'}^{(+)}\rangle
\langle {F}_{s, 3}^{(+)}{F}_{r, 4'}^{(+)}  \rangle,
\label{15}\end{equation}

\begin{equation}
\langle {F}_{DV1,x}^{(+)} {F}_{DH2,x}^{(+)} {F}_{DV3,y}^{(+)} {F}_{DH4,y}^{(+)}\rangle
=+\frac{1}{2} \langle {F'}_{p, 1'}^{(+)}{F'}_{q, 2}^{(+)}\rangle
\langle {F}_{p, 3'}^{(+)}{F}_{q, 4}^{(+)}  \rangle,
\label{16bb}
\end{equation}
\begin{equation}
\langle {F}_{DV1,x}^{(+)} {F}_{DV2,y}^{(+)} {F}_{DH3,y}^{(+)} {F}_{DH4,y}^{(+)}\rangle
=-\frac{1}{2} \langle {F'}_{p, 1'}^{(+)}{F'}_{q, 3}^{(+)}\rangle
\langle {F}_{p, 2'}^{(+)}{F}_{q, 4}^{(+)}  \rangle.
\label{16}\end{equation}

Let us note that there is a difference of signs in the two correlations
that result in a concrete joint detection in areas $2$ and $3$, as it
can be seen by comparing the equations \eqref{15bb} and \eqref{16bb},
or \eqref{15} and \eqref{16}. The same relation can be found between
the two correlations that result in a concrete joint detection in areas
$1$ and $4$, as it can be seen by comparing Eqs. \eqref{15bb} and
\eqref{15}, and also Eqs. \eqref{16bb} and \eqref{16}. Nevertheless,
there is no sign difference in the whole set of correlations given in
Eqs. \eqref{13} and \eqref{14}, in which two detectors of the same
area, $DH2$ and $DV2$, or $DH3$ and $DV3$, are involved (see Eqs.
(\ref{hyper0}) and (\ref{hyper3})). On the other hand, in both cases,
the detections concerning areas $1$ and $4$ correspond to orthogonal
polarizations. This is a key point in our treatment: the intrinsic
nature of entanglement swapping is related to the quadruple correlation
properties of the electromagnetic field, through the stochastic
properties of the zeropoint radiation. Hence, Eqs.
\eqref{13} and \eqref{14} represent the contribution of the addend
$(1/2)|\psi ^{+}\rangle_{14} |\psi ^{+}\rangle_{23}$ in Eq.
\eqref{40}, and Eqs.
\eqref{15bb} to \eqref{16} the contribution of the singlet states
$(1/2)|\psi ^{-}\rangle_{14}|\psi ^{-}\rangle_{23}$.

Let us emphasize that, although quadruple correlations can be obtained from the cross-correlations due to the Gaussian behaviour of the light field, there is no possibility to understand this phenomenon only by the consideration of Eqs. \eqref{22primera} to \eqref{22}, because these cross-correlations are associated with joint detections concerning a given detector at the BSM station, with another one of areas $1$ or $4$.

Let us now compute the four fold detection probabilities, for which we shall use Eq. (\ref{e10}). In each case, when we take into account all the values of the polarization indices \textit{$\lambda$, $\lambda'$, $\lambda''$} and \textit{$\lambda'''$}, $15$ addends are zero, precisely those ones that contain an amplitude coming from the zero point which enters the idle channel of the PBS. The only non zero term corresponds to one of the quadruple correlations given in equations \eqref{13} and \eqref{14} (\eqref{15bb} to \eqref{16}), in the case of the four probabilities $P_{DH1,DH2,DV2,DV4}$,  $P_{DH1,DH3,DV3,DV4}$, $P_{DV1,DH2,DV2,DH4}$ and $P_{DV1,DH3,DV3,DH4}$ ($P_{DH1,DH2,DV3,DV4}$,  $P_{DH1,DV2,DH3,DV4}$, $P_{DV1,DH2,DV3,DH4}$ and $P_{DV1,DV2,DH3,DH4}$).
For example, it can be easily shown that:

\begin{equation}
\frac{P_{{DV1,DHi,DVi,DH4}}}{{K_{{DV1\;}}K_{{DHi\;}}K_{{ DVi\;}} K_{{ DH4\;}}}}=\frac{1}{4}
|\langle {F'}_{p, 1'}^{(+)}{F'}_{q, i}^{(+)}\rangle|^2 |\langle
{F}_{p, i'}^{(+)}{F}_{q, 4}^{(+)}\rangle|^2\,\,\,;\,\,\,i=2, 3,
\label{koka}
\end{equation}
with similar expressions for the rest of the probabilities. Now, by considering the situation in which there is an identical distance from each source to the detectors, and the ideal situation of instantaneous four fold detection at a given time, we have

\begin{equation}
\frac{P_{{ DV1,DHi,DVi,DH4}}}{{K_{{DV1\;}}K_{{DHi\;}}K_{{ DVi\;}} K_{{ DH4\;}}}}=\frac{g^4|V|^4}{4}|\nu(0)|^4\,\,\,;\,\,\,i=2, 3.
\label{koka2}
\end{equation}
An identical result is obtained for the rest of the quadruple probabilities.

\item Now we shall analyse the eight quadruple correlations
corresponding to the situation in which the detections in areas $2$ and $3$ correspond to the same polarization.
Using (\ref{20}) and equations (\ref{22primera}) to (\ref{22}), there are six cuadruple correlations that vanish, those concerning three or four detectors with the same polarization. On the other hand, we have:
\[
\langle {F}_{DH1,x}^{(+)} {F}_{DV2,y}^{(+)}{F}_{DV3,y}^{(+)} {F}_{DH4,x}^{(+)} \rangle
\]
\begin{equation}
=\frac{-1}{2} \left[\langle  {F'}_{s, 1}^{(+)}{F'}_{r, 2'}^{(+)}
\rangle  \langle {F}_{p, 3'}^{(+)} {F}_{q, 4}^{(+)} \rangle
+i^{2}\langle {F'}_{s, 1}^{(+)} {F'}_{r, 3'}^{(+)} \rangle \langle
{F}_{p, 2'}^{(+)} {F}_{q, 4}^{(+)} \rangle\right].
\label{25}
\end{equation}

\[
\langle {F}_{DV1,y}^{(+)} {F}_{DH2,x}^{(+)}{F}_{DH3,x}^{(+)} {F}_{DV4,y}^{(+)} \rangle
\]
\begin{equation}
=\frac{-1}{2} \left[\langle  {F'}_{p, 1}^{(+)}{F'}_{q, 2'}^{(+)}
\rangle  \langle {F}_{s, 3}^{(+)} {F}_{r, 4'}^{(+)} \rangle
+i^{2}\langle {F'}_{p, 1'}^{(+)} {F'}_{q, 3}^{(+)} \rangle \langle
{F}_{s, 2}^{(+)} {F}_{r, 4'}^{(+)} \rangle\right]
.\label{25bis}\end{equation}

The factor $i^2$ that appears in equations (\ref{25}) and (\ref{25bis}) implies that, in the case in which there is an identical distance from the sources to the detectors, and we consider instantaneous four fold detection at a given time,
such correlations are null. In that case, there is a cancellation due to the action of the BS, through the factors $(+i^{2}/2)$ (two reflections) and
$1/2$ (two transmissions).


\item In the above situation, the two addens $-(1/2)| \Phi ^{+}\rangle_{14}| \Phi ^{+}  \rangle_{23}$
and $(1/2)| \Phi ^{-}\rangle_{14} |\Phi ^{-} \rangle_{23}$ cannot be
distinguished via single photon detectors, so that a double detection
occurs at a given detector placed in beam areas $2$ or $3$. In order to
analyse this possibility in the WRHP in terms of the quadruple
correlations, we shall study the situations that result in a double
screening in one of the detectors corresponding to areas $2$ or $3$.
This corresponds to the calculation of the four quadruple correlations
of the type $\langle {F}_{a}^{(+)} {F}_{b}^{(+)} {F}_{b}^{(+)}
{F}_{c}^{(+)}\rangle$, where $b \equiv \{DH2,x; DV2,y; DH3,x; DV3,y\}$,
and $a$ and $c$ are referred to detectors in areas $1$ and $4$
respectively, corresponding to the same polarization, and with the same
polarization component of the field, i.e. $(a, c)=(DH1,x; DH4, x)$ or
$(a, c)=(DV1,y; DV4, y)$. In this situation, if we consider the same
position and time for the ``double" detection at detector ``$b$", we
have:
\begin{equation}
\langle {F}_{a}^{(+)} {F}_{b}^{(+)} {F}_{b}^{(+)} {F}_{c}^{(+)}\rangle=2\langle {F}_{a}^{(+)} {F}_{b}^{(+)}\rangle
\langle {F}_{b}^{(+)} {F}_{c}^{(+)}\rangle.
\label{nnueva}
\end{equation}

By using Eq. \eqref{nnueva} and the cross-correlation properties given in Eqs. \eqref{22primera} to \eqref{22} we easily reach:

\begin{equation}
\langle {F}_{DV1,y}^{(+)} {F}_{DH2,x}^{(+)} {F}_{DH2,x}^{(+)} {F}_{DV4,y}^{(+)} \rangle
=i\langle {F'}_{p, 1'}^{(+)} {F'}_{q, 2}^{(+)} \rangle \langle
{F}_{s, 2}^{(+)}{F}_{r, 4'}^{(+)} \rangle,
\label{30}\end{equation}

\begin{equation}
\langle {F}_{DH1,x}^{(+)} {F}_{DV2,y}^{(+)} {F}_{DV2,y}^{(+)} {F}_{DH4,x}^{(+)} \rangle=
i\langle {F'}_{s, 1}^{(+)} {F'}_{r, 2'}^{(+)} \rangle \langle
{F}_{p, 2'}^{(+)} {F}_{q, 4}^{(+)}  \rangle, \label{31}\end{equation}

\begin{equation}
\langle {F}_{DV1,y}^{(+)} {F}_{DH3,x}^{(+)} {F}_{DH3,x}^{(+)} {F}_{DV4,y}^{(+)} \rangle=
i\langle {F'}_{p, 1'}^{(+)} {F'}_{q, 3}^{(+)}\rangle \langle
{F}_{s, 3 }^{(+)} {F}_{r, 4'}^{(+)} \rangle, \label{32}\end{equation}

\begin{equation}
\langle {F}_{DH1,x}^{(+)} {F}_{DV3,y}^{(+)} {F}_{DV3,y}^{(+)} {F}_{DH4,x}^{(+)} \rangle=
i\langle {F'}_{s, 1}^{(+)} {F'}_{r, 3'}^{(+)} \rangle \langle
{F}_{p, 3'}^{(+)} {F}_{q, 4}^{(+)}  \rangle. \label{33}\end{equation}

By inspection of Eqs. \eqref{30} to \eqref{33} we see that each of
these expressions correspond to the factorization of the two
cross-correlations that appear in Eqs. \eqref{22primera} to \eqref{22}
respectively. The information contained into the signs $+$ or $-$ in
Eqs. \eqref{22primera} to \eqref{22} is erased via this factorization.
This justifies that the addends $-(1/2)| \Phi ^{+}\rangle_{14}| \Phi
^{+}  \rangle_{23}$ and $(1/2)| \Phi ^{-}\rangle_{14} |\Phi ^{-}
\rangle_{23}$ cannot be distinguished.

\end{enumerate}

A quick look at Eqs. (\ref{13}) and (\ref{14}), and \eqref{15bb} to
\eqref{16} shows that each of these eight correlations is expressed in
terms of one of the two products: $\langle {F'}_{s}^{(+)}
{F'}_{r}^{(+)}
\rangle\langle {F}_{s}^{(+)} {F}_{r}^{(+)}
\rangle$ or $\langle {F'}_{p}^{(+)}
{F'}_{q}^{(+)}
\rangle\langle {F}_{p}^{(+)} {F}_{q}^{(+)}
\rangle$, evaluated at different positions and times. On the other hand,
by inspection of Eqs. (\ref{30}) to (\ref{33}) we see that each
correlation contains the product $\langle {F'}_{s}^{(+)}
{F'}_{r}^{(+)}
\rangle\langle {F}_{p}^{(+)} {F}_{q}^{(+)}
\rangle$ or $\langle {F'}_{p}^{(+)}
{F'}_{q}^{(+)}
\rangle\langle {F}_{s}^{(+)} {F}_{r}^{(+)}
\rangle$. These four products of cross-correlations have their origin
in the quadruple correlation properties of the beams outgoing the
crystals, given in Eqs. \eqref{GrindEQ__6_3_},
\eqref{GrindEQ__6_3_bis},
\eqref{GrindEQ__6_6_bis} and \eqref{GrindEQ__6_6_}, so that (for simplicity we discard
the dependence on position and time):

\begin{equation}
\langle {F}_{1,x}^{(+)} {F}_{2,y}^{(+)} {F}_{3,x}^{(+)} {F}_{4,y}^{(+)} \rangle
=\langle {F'}_{s}^{(+)}
{F'}_{r}^{(+)}
\rangle\langle {F}_{s}^{(+)} {F}_{r}^{(+)}
\rangle,
\label{love1}\end{equation}

\begin{equation}
\langle {F}_{1,x}^{(+)} {F}_{2,y}^{(+)} {F}_{3,y}^{(+)} {F}_{4,x}^{(+)} \rangle
=\langle {F'}_{s}^{(+)}
{F'}_{r}^{(+)}
\rangle\langle {F}_{p}^{(+)} {F}_{q}^{(+)}
\rangle,
\label{love2}\end{equation}

\begin{equation}
\langle {F}_{1,y}^{(+)} {F}_{2,x}^{(+)} {F}_{3,x}^{(+)} {F}_{4,y}^{(+)} \rangle
=\langle {F'}_{p}^{(+)}
{F'}_{q}^{(+)}
\rangle\langle {F}_{s}^{(+)} {F}_{r}^{(+)}
\rangle,
\label{love3}\end{equation}

\begin{equation}
\langle {F}_{1,y}^{(+)} {F}_{2,x}^{(+)} {F}_{3,y}^{(+)} {F}_{4,x}^{(+)} \rangle
=\langle {F'}_{p}^{(+)}
{F'}_{q}^{(+)}
\rangle\langle {F}_{p}^{(+)} {F}_{q}^{(+)}
\rangle,
\label{love4}\end{equation}
where we have used Eq. (\ref{12}) and considering that beams $1-2$ are
uncorrelated with beams $3-4$. The propagation of these quadruple
correlations through the experimental setup allows for the possibility
of transferring entanglement.

\section{Discussion and Conclussions}
\label{sec5}
Entangling particles that have never interacted is one of the most
interesting applications of entanglement to quantum information.
Nowadays, the idea of transmitting entanglement using the properties of
quantum correlations is a very important theoretical tool for the
development of quantum computer science. In this paper, the application
of the Wigner formalism to the theory of entanglement swapping with
photons generated via PDC opens a new framework for a deeper
understanding of this phenomenon and its applications to quantum
communication and conceptual problems of quantum mechanics. The WRHP
formalism gives a full quantum electrodynamical description of
entanglement teleportation, which contrasts to the usual particle-like
description using the qubit formalism and the striking application of
the projection postulate. The wavelike aspect of light is emphasized
throughout the role of the zeropoint field fluctuations in the
generation and measurement of quantum correlations.


We have applied the WRHP approach to analyse entanglement swapping,
first calculating the quadruple correlations of the field amplitudes
that characterize the horizontal and vertical components of the beams
$1$, $2'$, $3'$ and $4$ at the detectors, as functions of space-time
variables. From the analysis of these correlations, we can explain how,
although the cross-correlations between the pairs of beams $(1, 4)$,
$(2, 3)$ and $(2', 3')$ are zero, a Bell measurement on beams $2'$ and
$3'$ produces an entanglement swapping, according to the outcome of
Bell measurement on $2'$ and $3'$, to beams $1$ and $4$. In this way,
the correlation properties given in equations (\ref{13}) and (\ref{14})
are associated to the state $|\Psi^{+}_{23}\rangle
|\Psi^{+}_{14}\rangle$, in such a way that these four correlations keep
the same sign relations. In contrast, equations \eqref{15bb} to
\eqref{16} represent the four correlations that account for the state
$|\Psi^{-}_{23}\rangle |\Psi^{-}_{14}\rangle$, and the ``$+$" and
``$-$" signs that appear in these equations are closely related to the
intrinsic nature of the singlet state. Hence, the quadruple
correlations corresponding to the projection onto
$|\Psi^{+}_{23}\rangle |\Psi^{+}_{14}\rangle$ or $|\Psi^{-}_{23}\rangle
|\Psi^{-}_{14}\rangle$ characterize the exchange of properties between
the two couple of beams (1, 2) and (3, 4), to the couples (2, 3) and
(1, 4), which occurs in $50\%$ of cases. On the other hand, each of the
addends $-(1/2)|
\Phi ^{+}\rangle_{14}| \Phi ^{+}\rangle_{23}$ and $(1/2)|
\Phi ^{-}\rangle_{14} |\Phi ^{-}\rangle_{23}$
is represented by the same set of four correlations, just the
corresponding to Eqs. (\ref{30}) to (\ref{33}), which represent a
possible double detection in one of the detectors at the Bell state
analyser. For this reason, these states cannot be distinguished.

What is remarkable about our
formalism is that the modes involved in beams $1$ and
$4$ continue to be uncorrelated after the Bell measurement on
photons $2'$ and $3'$. In this way, these beams do not change in the ``non-local" form after a Bell measurement, which justifies the necessity of investigating the quadruple correlations of the light field.
This is a common feature to any experiment using the Wigner function of PDC, and
it contrasts to the analysis within the Hilbert space, where the collapse of the state vector, expressed in the Bell base of photons $2'$
and $3'$, gives rise to an entangled state in the
Hilbert space of photons $1$ and $4$.


Another crucial point of the WRHP of PDC is the
relationship between the zeropoint field inputs at the experimental setup
and the information that can
be obtained in a concrete experiment of quantum communication. This
possibility opens a way for a better understanding of quantum
communication using quantum optics. In \cite{swpJ} it was stressed that
two-photon entanglement in one degree of freedom (polarization) implies
the ``activation" of four independent sets of zeropoint modes at the
source, throughout a coupling with the laser inside the crystal.
In \cite{swpK} we have demonstrated that, for a given number of degrees
of freedom $n$, the maximal distinguishability in a Bell-like
experiment is bounded by the number of independent vacuum sets of modes
which are extracted at the source. The use of Hilbert
spaces of higher dimensions is related, within the WRHP approach, to
the inclusion of more sets of vacuum modes entering the source/s in
which the light is produced. With an increasing number of vacuum
inputs, the possibility for extracting more information from the
zeropoint field also increases, and also the capacity for using the
zeropoint amplitudes in quantum communication schemes.


Concretely, the generation of the product state of four photons (see
Eq. \eqref{40}) is represented, in the context of the WRHP approach,
via the consideration of eight sets of independent vacuum modes, four
corresponding to each pair emission, which are amplified for a further
registration at the detectors. Let us note that the eight sets of
amplified modes are included at the field amplitudes of beams $2'$ and
$3'$, as it can be seen from Eqs. \eqref{GrindEQ__6_3_bis} and
\eqref{GrindEQ__6_6_bis}. The beam-splitter does not introduce
additional zeropoint amplitudes, so that the information concerning the
eight sets of amplified modes enter the BSM analyser. There, the idle
channels of the PBSs constitute a fundamental input of noise in order
to ``brake" the beams $2'$ and $3'$ before the projective measurement.
Each idle channel introduces two sets of vacuum modes, each
corresponding to a given polarization, as it can be seen from Eqs.
\eqref{1000} to \eqref{61}. Let us note that the difference between the
total number of zeropoint sets of modes which are amplified at the two
crystals (eight) and the number of sets of vacuum modes entering the
idle channels at the BSM station (four), gives the four sets of vacuum
modes which are necessary for the description of an entangled pair of
photons. In this sense, the zeropoint field has a double role: on the
one hand, it is the ``carrier" of the quantum information which is
stored in the field amplitudes; on the other hand, the vacuum inputs at
the analyser introduce a fundamental noise giving rise to the
projective measurement. After the communication via classical
information, four sets of ``useful" amplified zeropoint amplitudes
remain, just the number for the description of entanglement in one
degree of freedom \cite{swpJ}.

Besides, the correlation properties of the electromagnetic field can be changed by
means of local operations in order to establish an entanglement swapping
teleportation protocol. For instance, taking into account that
equations (\ref{13}) and (\ref{14}) establish the correlation properties
corresponding to the projection onto the state $|\Psi^{+}_{23}\rangle
|\Psi^{+}_{14}\rangle$, a simple phase shift
$\alpha=\pi$ among vertical and horizontal components in beam
$4$, which produces the change $F_{r, 4'}^{(+)}\rightarrow
-F_{r, 4'}^{(+)}$, triggers a change of sign in equation (\ref{13}), being
reflected in the change $|\Psi^{+}_{23}\rangle
|\Psi^{+}_{14}\rangle\rightarrow |\Psi^{+}_{23}\rangle
|\Psi^{-}_{14}\rangle$. Victor (at the BSM station) must only inform Bob
about the result, and Bob would modify beam
$4$, in order to let beams
$1$ and $4$ entangled in the singlet state.
It is important to stress that this phase change acts directly on the
field amplitudes, so that the unitary operation performed by Bob, within the Hilbert-space
description, is closely linked to the modification of the properties of the ``amplified" vacuum,
through the action of optical devices operating in these experiments, resembling classical optics.

In general, the theoretical study of a given experimental arrangement of
entanglement swapping using the WRHP approach, for instance the ones
included in Refs. \cite{swpC}, \cite{swpB}, and \cite{nueva2013}, should take into account
Eqs. \eqref{GrindEQ__6_3_} to \eqref{GrindEQ__6_6_} corresponding to
the light beams outgoing the crystals, and the calculation of the
quadruple correlation properties of the electric field when the beams
are propagated from the source to the detectors. From the analysis of
these correlations, the use of the four fold detection probability
given in Eq. \eqref{e10}, and the study of the different zeropoint
entries at the experimental setup, this formalism can give a new
perspective to these experiments.
Concretely, the idea of transferring entanglement between two initially
non interacting particles is, not only an  important theoretical tool
for quantum computing, but also the starting point for the so called
delayed-choice entanglement swapping paradox \cite{swpZ}. Also, in a
recent paper it has been demonstrated that entanglement can be
generated between timelike separated quantum systems \cite{nueva2013}.
The WRHP approach allows for an explanation of these phenomena, in
which there is no quantum steering into the past, but a causal one
based on the correlation properties of the light field in terms of
space and time variables. For instance, in Ref. \cite{swpB} the
measurements performed by Alice and Bob are completely uncorrelated,
because the beams $1$ and $4$ do not share the same zeropoint
amplitudes. Nevertheless, the total information stored in the
electromagnetic field is finally extracted when Victor measures photons
$2$ and $3$, so that the classical communication of Victor's results to
Alice and Bob allows them to divide their results into subsets, which
can be used for Bell tests. On the other hand the idea of producing an
entanglement of a ``non-existing" particle with another one
\cite{nueva2013}, can be understood in a wave-like argument based on
the ZPF. In this way, the idea that photons are just an amplified
vacuum, and behave like waves until they are detected, is the key for
understanding that photon entanglement is just the evidence of the
possibility for manipulating the amplified vacuum, which is supported
by the quadruple correlations of the field. A deeper treatment of these
aspects will be made in further works.



\newpage

\appendix
\numberwithin{equation}{section}
\section{Appendix A: General aspects of the WRHP formalism}\label{A}
In this appendix, a brief review of the WRHP approach of PDC is
provided. The concepts and expressions presented in this appendix are
used extensively throughout this paper.

In the Heisenberg picture the electric field is represented by a
time-independent density operator corresponding to the initial state.
The electric field operator contains all the dynamics of the process,
through the time-dependent annihilation and creation operators
$\hat{a}_{{\bf k},
\lambda}(t)$ and $\hat{a}^{\dagger}_{{\bf k},
\lambda}(t)$. When passing to the Wigner representation, the
Wigner transformation establishes a correspondence between the field
operator and a time-dependent complex amplitude of the field, through
the substitution $\hat{a}_{{\bf k},
\lambda}(t)\rightarrow \alpha_{{\bf k}, \lambda}(t)$, $\hat{a}^{\dagger}_{{\bf k},
\lambda}(t)\rightarrow \alpha_{{\bf k}, \lambda}^{*}(t)$. The Wigner function is time-independent, and
corresponds to the Wigner distribution of the initial state.

In the context of PDC the initial state is the vacuum, being the Wigner
distribution for the vacuum field amplitudes \cite{swpL}:

\begin{equation}
W_{\it ZPF}(\{\alpha\})=
{\prod_{{\bf [k]}, \lambda}}\frac{2}{\pi}
{\rm e}^{-2|\alpha_{{\bf k}, \lambda}|^2},
\label{eq_w9}
\end{equation}
where $\alpha_{{\bf k}, \lambda}$ represents the zeropoint amplitude
corresponding to the mode $\{{\bf k}, \lambda\}$, and $\{\alpha\}$
represents the set of zeropoint amplitudes.

The electric field corresponding to a signal beam generated by the
nonlinear source (placed at ${\bf r}=0$) is represented by a slowly
varying amplitude \cite{swpM}:

\begin{equation}
{\bf F}^{(+)}({\bf r}, t)=i{\rm e}^{\omega_s
t}\sum_{{\bf k}\in [{\bf k}]_{s}, \lambda=H, V}\left(\frac{\hbar
\omega_{{\bf k}}}
{2\epsilon_0L^3}\right)^{\frac{1}{2}}\alpha_{{\bf k}, \lambda}(0){\bf
u}_{{\bf k}, \lambda}{\rm e}^{i({\bf k}\cdot{\bf r}-\omega_{\bf k}t)},
\label{F}
\end{equation}
where $[{\bf k}]_{s}$ represents a set of wave vectors centered at
${\bf k}_s$, and $\omega_s$ is the average frequency of the beam. ${\bf
u}_{{\bf k}, \lambda}$ is a unit polarization vector. On the other
hand, $\alpha_{{\bf k}, \lambda}(0)$ is a linear transformation, to
second order in the coupling constant ($g$) of the zeropoint field
entering the nonlinear crystal, which interacts with the laser beam
between $t=-\Delta t$ and $t=0$, $\Delta t$ being the interaction time.
For $t>0$ there is a free evolution.

The field amplitude ${\bf F}^{(+)}$ propagates through free space
according to the following expression \cite{swpL}:

\begin{equation}
{\bf F}^{(+)}({\bf{r}}_{2},t)=
{\bf F}^{(+)}({\bf{r}}_{1}, t-\frac{r_{12}}{c}){\rm e}^{i\omega_{s}\frac{r_{12}}{c}}
\,\,\,\,;\,\,\,\, {\bf{r}}_{12}=|{\bf{r}}_2-{\bf{r}}_1|.
\label{propagation}
\end{equation}

Given two complex amplitudes, $A({\bf r}, t; \{\alpha\})$ and $B({\bf
r'}, t';\{\alpha\})$, the correlation between them is given by:

\begin{equation}
\langle AB \rangle \equiv \int W_{\it ZPF}(\{\alpha\})A({\bf r}, t; \{\alpha\})B({\bf r'}, t';
\{\alpha\})d\{\alpha\}.
\label{corr}
\end{equation}
For instance, from (\ref{eq_w9}) the well known correlation properties
hold:

\begin{equation}
\langle \alpha_{{\bf k}, \lambda}\alpha_{{\bf k'}, \lambda'}\rangle
=\langle \alpha^*_{{\bf k}, \lambda}\alpha^*_{{\bf k'},
\lambda'}\rangle=0\,\,\,\,;\,\,\,\,\,\langle \alpha_{{\bf k}, \lambda}\alpha^*_{{\bf k'}, \lambda'}\rangle
=\frac{1}{2}\delta_{{\bf k}, {\bf k'}}\delta_{\lambda, \lambda'}.
\label{correlations}
\end{equation}

The single and joint detection probabilities in PDC experiments are
calculated, in the Wigner approach, by means of the expressions
\cite{swpM}:

\begin{equation}
P_{a}\propto \langle I_a-I_{ZPF, a} \rangle,
\label{probsimple}
\end{equation}

\begin{equation}
P_{ab}\propto
\langle(I_a-I_{ZPF, a})(I_b-I_{ZPF, b}) \rangle,
\label{prob}
\end{equation}
where $I_i\propto {\bf F}_{i}^{(+)}{\bf F}_{i}^{(-)}$, $i=a, b$, is the
intensity of light at the position of the $i$-detector, and $I_{ZPF,
i}$ is the corresponding intensity of the zeropoint field. In actual
experiments the expressions given by (\ref{probsimple}) and
(\ref{prob}) must be integrated over appropriate detection windows and
the surface of the detectors.

In experiments involving polarization, the following simplified
expression for the joint detection probability will be used for
practical matters:

\begin{equation}
P_{ab} \left({\bf{r}},t;{\bf{r'}},t'\right)\propto
\sum _{\lambda, \lambda'}
\left|\left\langle F_{a, {\lambda }}^{\left(+\right)} \left(\phi_{a} ;{\bf{r}},t\right)
F_{b, {\lambda '}}^{\left(+\right)} \left(\phi_{b} ;{\bf{r'}},
t'\right)\right\rangle \right|^{2},
\label{p12}
\end{equation}
where and $\phi_{A}$ and $\phi_{B}$ are controllable parameters of the
experimental setup.

In experiments involving two pairs of photons emitted by independent
sources, the quadruple correlation between four complex amplitudes
$A({\bf{r}}, t;
\alpha)$, $B({\bf{r}}', t';
\alpha)$, $C({\bf{r}}'', t''; \alpha)$ and $D({\bf{r}}''', t''';
\alpha)$, is given by:

\begin{equation}
\langle ABCD \rangle \equiv \int W_{ZPF}(\alpha)A({\bf{r}}, t; \alpha)
B({\bf{r}}', t'; \alpha)C({\bf{r}}'', t''; \alpha)D({\bf{r}}''', t''';
\alpha) d\alpha.
\label{pepi1}
\end{equation}

In the case of PDC light, and taking into account that we are dealing
with a Gaussian process, each quadruple correlation is expressed in
terms of double correlations as:

\begin{equation}\begin{array}{l} {\left\langle ABCD \right\rangle=
\left\langle AB \right\rangle \left\langle CD \right\rangle +}\left\langle AC \right\rangle
\left\langle BD \right\rangle  +\left\langle AD \right\rangle  \left\langle BC \right\rangle.
\end{array}\label{12}\end{equation}

Finally, the quadruple detection probabilities, which are necessary in
experiments on teleportation, are given by (see Appendix \ref{B}):

\[
P_{abcd}({\bf{r}}, t; {\bf{r}}', t'; {\bf{r}}'', t''; {\bf{r}}''',
t''')
\]
\begin{equation}
\propto \sum_{\lambda, \lambda', \lambda'', \lambda'''}|\langle {F}_{a,\lambda }^{(+)} (\phi_{a} ; {\bf{r}}, t)
{F}_{b,\lambda {'} }^{(+)} (\phi_{b} ; {\bf{r}}', t')
{F}_{c,\lambda {'} {'} }^{(+)} (\phi_{c} ; {\bf{r}}'', t''){F}_{d,\lambda {'} {'} {'} }^{(+)}
(\phi_{d} ; {\bf{r}}''', t''')\rangle|^{2}.
\label{e10}
\end{equation}

The role of the zeropoint field as a threshold for detection can be put
explicitly, by taking into account that Eq. \eqref{e10} coincides with
the following expression, for PDC experiments:

\begin{equation}  \begin{array}{l} {P_{abcd} \propto
\langle \left(I_{a} -I_{ZPF,a} \right)\left(I_{b} -I_{ZPF,b} \right)\left(I_{c} -I_{ZPF,c} \right)
\left(I_{d} -I_{ZPF,d} \right)\rangle. } \\  \end{array} \end{equation}

A key point of the WRHP formalism of PDC is the description of
entanglement. In this approach, entanglement appears just as an
interplay of correlated waves, through the distribution of the vacuum
amplitudes in the different polarization components of the field
\cite{swpM}. For instance, quantum predictions corresponding to the
states $|\Psi^\pm\rangle$ are reproduced in the Wigner framework by
considering the following two correlated beams outgoing the crystal
\cite{swpJ}:

\begin{equation}
{\bf F}_{1}^{(+)}({\bf r}, t)=F_s^{(+)}({\bf r},
t; \{\alpha_{{\bf k}_1, H}; \alpha^*_{{\bf k}_2, V}\}){\bf
i}+F_{p}^{(+)}({\bf r}, t;
\{\alpha_{{\bf k}_1, V}; \alpha^*_{{\bf k}_2, H}\}){\bf j}
\label{hyper0},
\end{equation}
\begin{equation}
{\bf F}_{2}^{(+)}({\bf r}, t)=F_{q}^{(+)}({\bf r},
t; \{\alpha_{{\bf k}_2, H}; \alpha^*_{{\bf k}_1, V}\}){\bf
i'}\pm F_{r}^{(+)}({\bf r}, t;
\{\alpha_{{\bf k}_2, V}; \alpha^*_{{\bf k}_1, H}\}){\bf j'},
\label{hyper3}
\end{equation}
where ${\bf i}$ and ${\bf i'}$ (${\bf j}$ and ${\bf j'}$) are unit
vectors representing horizontal (vertical) linear polarization at beams
``$1$" and ``$2$", and $\{\alpha_{{\bf k}_i, V};\alpha_{{\bf k}_i,
H}\}$\,($i=1, 2$) represent four sets of relevant zeropoint amplitudes
entering the crystal. The four set of modes $\{{\bf k}_{i,
\lambda}\}$\,($i=1, 2$;  $\lambda\equiv H, V$) are ``activated" and
coupled with the laser beam inside the nonlinear medium. In expressions
(\ref{hyper0}) and (\ref{hyper3}), the only non vanishing correlations
are those involving the combinations
$r
\longleftrightarrow s$ and $p \longleftrightarrow q$. Hence, the non-null cross-correlations
correspond to different polarization components, the only difference being the minus sign that
appears in the case of $|\Psi^{-}\rangle$.
These correlations are directly related to the way in which the vacuum
components are distributed inside the total field amplitudes (see Eq.
\ref{correlations}).

By using Eq. (\ref{propagation}) the cross-correlations, at any
position and time, can be expressed in terms of the corresponding ones
at the center of the nonlinear source \cite{swpM}. We have:

\begin{equation}
\langle F^{(+)}_{p}({\bf 0}, t)F^{(+)}_{q}({\bf 0}, t')\rangle =gV\nu(t'-t),
\label{nu}
\end{equation}
where $V$ is the amplitude of the laser beam. $\nu(t'-t)$ is a function
which vanishes when $|t'-t|$ is greater than the correlation time
between the amplitudes $F^{(+)}_{p}$ and $F^{(+)}_{q}$ \cite{swp19}.
Similar expression holds for $\langle F^{(+)}_{r}({\bf 0},
t)F^{(+)}_{s}({\bf 0}, t')\rangle$.

Finally, the beams corresponding to the states $|\Phi^\pm\rangle$ are:

\begin{equation}
{\bf F}_{1}^{(+)}({\bf r}, t)=F_s^{(+)}({\bf r},
t; \{\alpha_{{\bf k}_1, H}; \alpha^*_{{\bf k}_2, V}\}){\bf
i}+F_{p}^{(+)}({\bf r}, t;
\{\alpha_{{\bf k}_1, V}; \alpha^*_{{\bf k}_2, H}\}){\bf j}
\label{hyper000},
\end{equation}
\begin{equation}
{\bf F}_{2}^{(+)}({\bf r}, t)=F_{r}^{(+)}({\bf r}, t;
\{\alpha_{{\bf k}_2, V}; \alpha^*_{{\bf k}_1, H}\}){\bf
i'}\pm F_{q}^{(+)}({\bf r},
t; \{\alpha_{{\bf k}_2, H}; \alpha^*_{{\bf k}_1, V}\}){\bf j'}.
\label{hyper333}
\end{equation}
In this case, the non vanishing correlations correspond to the same polarization components,
the only difference being the minus sign that
appears in the case of $|\Phi^{-}\rangle$.

\newpage

\section{Appendix B: The quadruple detection probability in the WHRP}
\label{B}
The four-fold detection probability is usually expressed, in the
Hilbert space, by means of the following expectation value of a
normally ordered expression of electric field operators:

\begin{equation}
P_{abcd}=K_aK_bK_cK_d\sum_{\lambda, \lambda', \lambda'',
\lambda'''}
\langle \hat{F}_{a, \lambda}^{(-)} \hat{F}_{b, \lambda'}^{(-)}
\hat{F}_{c, \lambda''}^{(-)}\hat{F}_{d, \lambda'''}^{(-)}
\hat{F}_{d, \lambda'''}^{(+)} \hat{F}_{c, \lambda''}^{(+)}
\hat{F}_{b, \lambda'}^{(+)} \hat{F}_{a, \lambda}^{(+)}\rangle ,
\label{8}
\end{equation}
where $\lambda$, $\lambda'$, $\lambda''$ and $\lambda'''$ are
polarization indices, and $K_a$, $K_b$, $K_c$ and $K_d$ are constants
related to the detection efficiency. For simplicity, we shall use the
following notation

\begin{equation}
\hat{F}_{\alpha , \nu }^{(-)} \equiv \alpha\,\,\,;\,\,\,\hat{F}_{\alpha, \nu}^{(+)}\equiv \alpha'\,\,\,,\,\,\,
\alpha=\{a, b, c, d\}\,\,\,,\,\,\,\nu=\{\lambda, \lambda', \lambda'',
\lambda'''\},
\end{equation}
so that the four-fold detection probability is expressed by means of
the average $\langle a b c d d' c' b' a'\rangle$. By using the Wick's
theorem \cite{Tele5} we should have to consider, in principle, $105$
addends, each of them consisting on the product of four
cross-correlations. In case the correlation $\langle\hat{F}_{\alpha,
\lambda}^{(+)}\hat{F}_{\beta, \lambda'}^{(-)}\rangle$
($\langle\hat{F}_{\alpha, \lambda}^{(+)}\hat{F}_{\beta,
\lambda'}^{(+)}\rangle$) is not null,
it is of order $g^2$ ($g$) in PDC experiments. By retaining up to
fourth order terms in $g$, we have:

\begin{equation}  \begin{array}{l} {\langle a b c d d' c' b' a'\rangle}\\
{=\langle ab \rangle \left[\left\langle cd\right\rangle
\left(\left\langle d{'} c{'} \right\rangle \left\langle b{'} a{'} \right\rangle
+\left\langle d{'} b{'} \right\rangle \left\langle c{'} a{'}
\right\rangle +\left\langle d{'} a{'} \right\rangle
\left\langle c{'} b{'} \right\rangle \right)\right. }  {\left. \right]}
\\ {+\left\langle ac\right\rangle \left[\left\langle bd\right\rangle
\left(\left\langle d{'} c{'} \right\rangle \left\langle b{'} a{'} \right\rangle
+\left\langle d{'} b{'} \right\rangle \left\langle c{'} a{'}
\right\rangle +\left\langle d{'} a{'} \right\rangle \left\langle c{'}
b{'} \right\rangle \right)\right. }{\left. \right]} \\
{{\kern 1pt} +\left\langle ad\right\rangle \left[\left\langle bc\right\rangle \left(\left\langle d{'} c{'}
\right\rangle \left\langle b{'} a{'} \right\rangle
+\left\langle d{'} b{'} \right\rangle \left\langle c{'} a{'}
\right\rangle +\left\langle d{'} a{'}
\right\rangle \left\langle c{'} b{'} \right\rangle \right)\right. } {\left. \right]} {\kern 1pt}
.\end{array}
\label{7}
\end{equation}

In order to go to the Wigner representation, we shall use the fact that
the field operators $\hat{F}_{\alpha ,
\lambda }^{(+)}$ and $\hat{F}_{\beta,
\lambda'}^{(+)}$ commute, so that the following relation holds
\begin{equation}  \langle \hat{F}_{\alpha , \lambda }^{(+)} \hat{F}_{\beta, \lambda'}^{(+)}\rangle
=\langle S\left(\hat{F}_{\alpha, \lambda}^{(+)}\hat{F}_{\beta,
\lambda'}^{(+)} \right)\rangle =\langle {F}_{\alpha ,\lambda
}^{(+)}{F}_{\beta, \lambda'}^{(+)}\rangle_{W}, \label{5}\end{equation}
where $S()$ means symmetrization \cite{swpL}, and $\langle \rangle_{W}$
represents an average with the Wigner function of the quantum state of
the electromagnetic field. After some easy algebra, we arrive to the
following expression:

\begin{equation}\begin{array}{l}
{P_{abcd}=K_{a}K_{b}K_{c}K_{d}
\left[P_{ab} P_{cd} +
{\kern 1pt} P_{ac} P_{bd}+P_{ad} P_{bc}  \right.} \\
{+\sum_{\lambda, \lambda', \lambda'', \lambda'''}
\left(\left\langle {F}_{a,\lambda }^{(-)}
{F}_{b,\lambda {'} }^{(-)} \right\rangle _{W}
\left\langle {F}_{c,\lambda {'} {'} }^{(-)} {F}_{d,\lambda {'} {'} }^{(-)}
\right\rangle _{W} \left\langle {F}_{b,\lambda {'} }^{(+)} {F}_{d,\lambda {'} {'} {'} }^{(+)}
\right\rangle _{W} \left\langle {F}_{a,\lambda }^{(+)} {F}_{c,\lambda {'} {'} }^{(+)}
\right\rangle _{W} \right. } \\ {+\left\langle {F}_{a,\lambda }^{(-)} {F}_{b,\lambda {'} {'} }^{(-)}
\right\rangle _{W} \left\langle {F}_{c,\lambda {'} {'} }^{(-)} {F}_{d,\lambda {'} {'} {'} }^{(-)}
\right\rangle _{W} \left\langle {F}_{d,\lambda {'} {'} {'} }^{(+)} {F}_{a,\lambda }^{(+)}
\right\rangle _{W} \left\langle {F}_{c,\lambda {'} {'} }^{(+)} {F}_{b,\lambda {'} }^{(+)}
\right\rangle _{W} } \\ {+\left.\langle {F}_{a,\lambda }^{(-)} {F}_{c,\lambda {'} {'} }^{(-)}
\right\rangle _{W} \left\langle {F}_{b,\lambda {'} }^{(-)} {F}_{d,\lambda {'} {'} {'} }^{(-)}
\right\rangle _{W} \left\langle {F}_{d,\lambda {'} {'} {'} }^{(+)} {F}_{a,\lambda }^{(+)}
\right\rangle _{W} \left\langle {F}_{b,\lambda {'} }^{(+)} {F}_{c,\lambda {'} {'} }^{(+)}
\right\rangle_{W}} {\left.\left. + c.c. \right)\right].} \end{array}\label{9}\end{equation}

A deeper inspection of \eqref{9} gives us the final expression:

\begin{equation}
P_{abcd}=K_aK_bK_cK_d
\sum_{\lambda, \lambda', \lambda'', \lambda'''}
\left|\langle {F}_{a,\lambda }^{(+)} {F}_{b,\lambda {'} }^{(+)} {F}_{c,\lambda {'} {'} }^{(+)}
{F}_{d,\lambda {'} {'} {'} }^{(+)}\rangle_{W} \right|^{2}.
\label{e10e}\end{equation}

\end{document}